\documentclass[12pt,a4paper]{article}

\usepackage[left=2cm,right=2cm,top=2.5cm,bottom=2.5cm]{geometry}
\usepackage[utf8]{inputenc}

\usepackage{scalerel,stackengine,amsmath}
\usepackage{amsfonts}
\usepackage{amssymb}
\usepackage{hyphenat}
\usepackage{enumitem}
\usepackage{dsfont}
\usepackage{cite}
\usepackage{hyperref}
\usepackage{graphicx}
\usepackage{caption}
\usepackage{float}
\usepackage{subcaption}
\usepackage{soul}
\usepackage[labelformat=simple]{subcaption}
\usepackage{array}
\usepackage{physics}
\usepackage{float}
\usepackage{mathtools}
\usepackage{rotating}
\usepackage{amssymb}

\newcommand{\citet}{\cite} 
\usepackage{braket}
\usepackage[export]{adjustbox}
\usepackage{mathtools}
\usepackage{comment}
\usepackage{mathrsfs}  

\usepackage{adjustbox}
\usepackage{slashed}
\usepackage{mathrsfs}
\usepackage{lmodern}
\usepackage{bbold}
\usepackage{mmacells}

\numberwithin{equation}{section}
\renewcommand{\thefootnote}{\fnsymbol{footnote}}

\begin{document}

\begin{flushright}
October 2024
\end{flushright}

\begin{center}
{\bf \LARGE{\hspace{-4mm}Dirac Algebra Formalism for
                         Two Higgs Doublet\\[3mm] \hspace{-2mm}Models: the
                         One-Loop Effective Potential} }
\end{center}

\bigskip

\begin{center}
{\large Apostolos Pilaftsis}$\,$\footnote{E-mail address: {\tt
apostolos.pilaftsis@manchester.ac.uk}}\\[3mm] 
{\it Department of Physics
and Astronomy, University of
Manchester,\\ Manchester M13 9PL, United Kingdom}
\end{center}

\bigskip

\centerline{\bf ABSTRACT}
\vspace{2mm}

\noindent
We present a novel covariant bilinear formalism for the Two Higgs Doublet Model (2HDM) which utilises the Dirac algebra associated with the $\text{SL}(2,\mathbb{C})$ group that acts on the scalar doublet field space. This Dirac-algebra approach enables us to obtain a fully $\text{O}(1,3)$-covariant and IR-safe expression for the one-loop effective potential. We illustrate how the formalism can be used to evaluate the breaking of global symmetries of the 2HDM potential by loop effects, in a 
field-reparameterisation invariant manner.


\thispagestyle{empty}

\section{Introduction}\label{sec:intro}

The addition of one Higgs doublet to the field content of the Standard Model (SM) engenders\- a well-founded theoretical framework, known as the Two Higgs Doublet Model~(2HDM)~\cite{Lee:1973iz}, which allows us to address two longstanding cosmological problems in the Universe. First, unlike the SM, the 2HDM can account for the Dark-Matter problem if the extra scalar doublet is stable~\cite{Silveira:1985rk}. Second, it can provide  new sources of  CP violation of  spontaneous~\cite{Lee:1973iz,Branco:1980sz,Branco:1985aq},  explicit~\cite{Weinberg:1990me,Pilaftsis:1999qt} or even mixed origin~\cite{Darvishi:2023fjh}. Therefore, in contrast to the SM~\cite{Farakos:1994kx,Gavela:1994dt}, these new sources can give rise   to  electroweak baryogenesis~\cite{Kuzmin:1985mm}, through  a strong first  order phase transition~\cite{Cohen:1993nk}. 

In addition to  the electroweak gauge group SU(2)$_L\times$U(1)$_Y$, the SM scalar potential possesses one additional global (accidental) symmetry in the unbroken phase of the theory. This~symmetry is  called {\em custodial symmetry}~\cite{Sikivie:1980hm},
usually denoted as SU(2)$_C$, and manifests itself in the limit of a vanishing hypercharge coupling~$g'$ of U(1)$_Y$ and when the up- and down-quark Yukawa-coupling matrices, $h^u$ and $h^d$, are equal. The custodial symmetry leaves the SM-scalar potential 
invariant under SU(2)$_C$ transformations between the Higgs doublet~$\phi$ and its hypercharge-conjugate counterpart, $\widetilde{\phi} \equiv i\sigma^2\phi^*$. It also leaves invariant the kinetic term of the Higgs doublet, whilst being gauged under the electroweak SU(2)$_L$ group only.

As opposed to the SM~\cite{Paschos:1976ay,Sikivie:1980hm}, the 2HDM could realise  a wealth of  global  symmetries~\cite{Peccei:1977hh,Deshpande:1977rw,Ivanov:2007de,Battye:2011jj,Pilaftsis:2011ed}, continuous or discrete, whose breaking may  result  in  pseudo-Goldstone bosons~\cite{Weinberg:1972fn},  mass hierarchies,  flavour-changing  neutral currents~\cite{Glashow:1976nt, Pich:2009sp}
and  CP violation~\cite{Branco:1980sz,Branco:1985aq,Botella:1994cs,Darvishi:2023fjh}.
Specifically, the tree-level 2HDM potential may have 6 symmetries that preserve U(1)$_Y$~\cite{Ivanov:2007de}, as well as 7 extra symmetries that are custodial and occur when $g' \to 0$~\cite{Battye:2011jj,Pilaftsis:2011ed}, even though several of these global symmetries may require the absence of the Yukawa couplings, so as to be extended to the complete 2HDM Lagrangian. Remarkably enough, the spontaneous symmetry breaking of the continuous and discrete symmetries can give rise to topological field configurations~\cite{Battye:2011jj}, such as domain walls, vortices and global monopoles, which may leave imprints in the observable Universe~\cite{Battye:2020sxy,Battye:2020jeu,Law:2021ing,Eto:2021dca,Sassi:2023cqp}. On the other hand, several studies have been devoted to analyse the pheno\-menological impact of these symmetries~\cite{Ginzburg:2004vp,Branco:2005em,Davidson:2005cw,Gunion:2005ja}, including the exciting possibility of {\em natural alignment}~\cite{BhupalDev:2014bir,Pilaftsis:2016erj,Aiko:2020atr,
Darvishi:2023fjh}  between the gauge and Yukawa couplings of a light neutral 2HDM scalar and the respective couplings of the SM Higgs boson.

To unambiguously identify all the global symmetries of the 2HDM potential, one must utilise the so-called covariant bilinear formalism, introduced in~\cite{Maniatis:2006fs,Nishi:2006tg,Ivanov:2006yq}. In this formalism, one  maps covariantly the four scalar-doublet bilinears, $\phi^\dagger_i\phi_j$ (with $i,j=1,2$), to the four-vector~\cite{Ivanov:2007de}:
$r^\mu \equiv \mbox{\boldmath $\phi$}^\dagger \sigma^\mu 
\mbox{\boldmath $\phi$}$, where 
$\mbox{\boldmath $\phi$} = (\phi_1\,, \phi_2)^{\sf T}$ and $\sigma^\mu = ({\bf 1}_2\,, \mbox{\boldmath $\sigma$})$ is the Pauli four-vector. Hence, SL(2,$\mathbb{C})$ or SU(2) reparameterisations of the field-space vector $\mbox{\boldmath $\phi$}$ get translated into~O(1,3) or SO(3) rotations of the four-vector $r^\mu$ in the bilinear field space. In this bilinear field-space, the 2HDM potential can be written in a covariant quadratic form in powers of $r^\mu$, by means of which it becomes easier to identify the 6 U(1)$_Y$-invariant global symmetries as proper and improper subgroups of SO(3)~\cite{Ivanov:2007de,Ferreira:2009wh}. However, to identify the remaining 7 custodial symmetries, the above formalism requires a non-trivial extension~\cite{Battye:2011jj,Pilaftsis:2011ed}, where the field space needs to be enlarged to ${\bf \Phi}$ that includes the hypercharge-conjugate fields $\widetilde{\phi}_i = i\sigma^2 \phi^*_i$, as we will see in more detail in Section~\ref{sec:Dirac}. 

Going beyond the classical approximation, the efficiency  of the covariant bilinear formalism and its practicality have not been  investigated so far in adequate detail. Although encouraging multi-loop calculations have  appeared in the recent literature~\cite{Bednyakov:2018cmx} that include the one-loop effective potential~\cite{Cao:2022rgh,Cao:2023kgq},
a fully O(1,3)-covariant approach that takes into account\- the quantum effects from the complete 2HDM Lagrangian is still lacking. The existence of such an approach would be useful to better examine the origin of the Renormalization Group~(RG) invariants observed in~\cite{Ferreira:2023dke}, or other related methods based on {\em reduction equations} of the 2HDM~\cite{Pech:2023bjm}. Moreover, 
going beyond the Born approximation will provide an intuitive information\- about the breaking of symmetries in classically scale-invariant 2HDMs~\cite{Lee:2012jn,Eichten:2022vys}, or even allow to determine with greater precision the evolution decay rates of topological defects, caused by symmetry-violating terms induced at loop level.  

In this paper, we will present a new formulation of the covariant bilinear formalism for the Two Higgs Doublet Model (2HDM). Unlike earlier considerations, the new formulation relies on the Dirac algebra associated with the $\text{SL}(2,\mathbb{C})$ group that acts on the scalar doublet field space. This new Dirac-algebra approach will allow us to efficiently carry out covariant computations in the bilinear space, and as such, obtain
a fully $\text{O}(1,3)$-covariant and Infra-Red (IR)-safe expression for the one-loop effective potential. We will show how the formalism can be used to evaluate the breaking of global symmetries of the 2HDM potential by loop effects, in a field-reparameterisation invariant manner. As archetypal example for our illustrations, we will consider the so-called Maximally Symmetric Two Higgs Doublet Model~(MS-2HDM) which realises the maximal custodial symmetry group: Sp(4)~\cite{BhupalDev:2014bir,Pilaftsis:2016erj,Darvishi:2019ltl}. 

The remainder of the paper is organised as follows. In Section~\ref{sec:Dirac}, we  outline all key features of the Dirac-algebra formalism, when applying it to the 2HDM Lagrangian. Besides reviewing the 2HDM potential in the bilinear formalism, we discuss the new covariant form describing the kinetic terms of the scalar doublets, along with the Yukawa sector. In particular, we emphasise the role of the wave-function four-vector $\zeta_\mu$ in obtaining manifestly covariant results in the O(1,3) bilinear space. In Section~\ref{sec:effpot}, we compute the one-loop effects from the scalar-doublets, the gauge bosons and the fermions on the effective potential. In particular, we discuss the O(1,3)-covariant structure of these effects and their impact on the MS-2HDM. Section~\ref{sec:concl} contains our conclusions and summarises the important findings of the present study.

\renewcommand{\thefootnote}{\arabic{footnote}}
\section{Dirac Algebra Formalism for the 2HDM}\label{sec:Dirac}

Before describing our Dirac  field-space formalism, let us
first write down the relevant part of the 2HDM Lagrangian using
standard conventions,
\begin{equation}
   \label{eq:L2HDM}
  {\cal L}\;  =\; (D_\alpha\phi_1)^\dagger (D^\alpha \phi_1)\, +\,
  (D_\alpha\phi_2)^\dagger (D^\alpha \phi_2)\, -\, V\, +\, {\cal L}_{\rm Y}\; . 
\end{equation}
Here, $D_\alpha= {\bf 1}_2\partial_\alpha + \frac{i}{2}g \sigma^i W^i_\alpha
+\frac{i}{2} g' {\bf 1}_2B_\alpha$ is the covariant spacetime
derivative with respect to the SM gauge group that acts on the two
Higgs doublets~$\phi_1$ and~$\phi_2$ (with $\sigma^{i\,=\,1,2,3}$ denoting
the three Pauli matrices), and~$V$ is the tree-level scalar potential,
\begin{eqnarray}
  \label{eq:V2HDM}
V \!& = &\! m_{11}^2 (\phi_1^{\dagger} \phi_1) + m_{22}^2
(\phi_2^{\dagger} \phi_2) + m_{12}^2 (\phi_1^{\dagger} \phi_2) +
m_{12}^{*2}(\phi_2^{\dagger} \phi_1)
+ \lambda_1 (\phi_1^{\dagger} \phi_1)^2 + \lambda_2 (\phi_2^{\dagger} \phi_2)^2\nonumber\\
\!&&\! +\: \lambda_3 (\phi_1^{\dagger}
\phi_1)(\phi_2^{\dagger} \phi_2) + \lambda_4 (\phi_1^{\dagger}
\phi_2)(\phi_2^{\dagger} \phi_1)  + \frac{\lambda_5}{2} (\phi_1^{\dagger} \phi_2)^2 +
\frac{\lambda_5^{*}}{2} (\phi_2^{\dagger} \phi_1)^2\\\
\!&&\! +\: \lambda_6 (\phi_1^{\dagger} \phi_1) (\phi_1^{\dagger} \phi_2) + \lambda_6^{*}
(\phi_1^{\dagger} \phi_1)(\phi_2^{\dagger} \phi_1) + 
\lambda_7 (\phi_2^{\dagger} \phi_2) (\phi_1^{\dagger} \phi_2) +
\lambda_7^{*} (\phi_2^{\dagger} \phi_2) (\phi_2^{\dagger} \phi_1)\; .\nonumber
\end{eqnarray}
The general CP-violating 2HDM potential $V$~\cite{Pilaftsis:1999qt} contains 4 real mass parameters,
$m_{11}^2$, $m_{22}^2$, ${\rm Re}(m_{12}^2)$ and ${\rm Im}(m^2_{12})$,
and 10 real quartic couplings, $\lambda_{1,2,3,4}$,
${\rm Re}(\lambda_{5,6,7})$ and ${\rm Im}(\lambda_{5,6,7})$.  Note~that all these 14~parameters are required for the renormalisability of the theory, in its off-shell formulation~\cite{Weinberg:1990me} that includes the effective potential~\cite{Pilaftsis:1997dr,Pilaftsis:1998pe}.
Finally, the last term ${\cal L_{\rm Y}}$ on the RHS of~\eqref{eq:L2HDM}
describes the Yukawa interactions of the Higgs doublets to quarks and leptons.

An equivalent $\text{O}(1,3)$-covariant description of the 2HDM
Lagrangian may be obtained by adopting a {\em variant} of the 8D
representation introduced in~\cite{Battye:2011jj,Nishi:2011gc,Pilaftsis:2011ed}, which
allows to collectively account for all scalar fields in the theory by the
multiplet,
\begin{equation}
 \label{eq:Phi}
 {\bf \Phi}\ \equiv\ \{\Phi^a\}\: =\:
 \Big( {\bf 1}_2\oplus i\sigma^2\Big) \otimes {\bf 1}_2\, \left( \begin {array}{c}
                                       \phi_1\\[2mm]
                                       \phi_2\\[2mm]
                                       i \sigma^2 \phi_1^*\\[2mm]
                                      i \sigma^2 \phi_2^* \end {array}
                                  \right)\: =\: 
 \left(\!\!\begin {array}{c}
                   \ \phi_1\\[2mm]
                   \ \phi_2\\[2mm]
\big(i\sigma^2\otimes {\bf 1}_2\big)\left(\!\begin{array}{c}
                   i \sigma^2 \phi_1^*\\[2mm]
        i \sigma^2 \phi_2^*
\end{array}\!\right)
\end {array}\!\!\right)
\,.            
\end{equation}
Here, a lowercase Latin letter, such as
$a,b = 1,2,3,4$, labels the four Higgs-doublet elements contained
in the ${\bf \Phi}$ multiplet. In this respect, we observe that each
doublet component $\Phi^a$ of ${\bf \Phi}$ transforms multiplicatively
under an ${\rm SU}(2)_L$ gauge rotation~${\rm U}_L$:
$\Phi^a \to \Phi'^a = {\rm U}_L\, \Phi^a$, and hence
${\bf \Phi}' = {\rm U}_L {\bf \Phi}$. Instead, under the SM
hypercharge group, ${\rm U(1)}_Y$, the two upper and two lower
scalar-doublet elements of ${\bf \Phi}$ transform differently:
$\Phi'^{1,2} = e^{i\theta}\, \Phi^{1,2}$ and
$\Phi'^{3,4} = e^{-i\theta}\, \Phi^{3,4}$, with
$e^{i\theta}\in {\rm U(1)}_Y$.  Because of these different
transformation properties under ${\rm U(1)}_Y$, the multiplet
${\bf \Phi} =\{\Phi^a\}$ (with $a=1,2,3,4$) can be identified as a
four-component Majorana fermion (see our discussion below in~\eqref{eq:MajoranaPhi}), which features the following left- and
right-handed `chiral' decomposition:
\begin{equation}
    \label{eq:PhiLR}
 {\bf \Phi}\: =\: {\bf \Phi}_L\, +\, {\bf \Phi}_R\,,
\end{equation}
where ${\bf \Phi}_L = \{ \Phi^1\,, \Phi^2\,, 0\,, 0\}$ and
${\bf \Phi}_R = \{ 0\,, 0\,, \Phi^3\,, \Phi^4\}$.

It is now convenient to introduce the generators,
$\sigma^\mu = ({\bf 1}_2\,, \mbox{\boldmath $\sigma$} )$ and
$\bar{\sigma}^\mu = ({\bf 1}_2\,, -\mbox{\boldmath $\sigma$} )$, with $\sigma^0 = {\bf 1}_2$ and~\mbox{\boldmath
$\sigma$}$= \{\sigma^{1,2,3}\}$, for the two conjugate spinorial 2D Clifford algebras associated with the ${\rm SL}(2\,,\mathbb{C})$ group~\cite{Garling:2011,Wess:1992}:
\begin{equation}
  \label{eq:Clifford}
 \sigma^\mu \bar{\sigma}^\nu \, +\, \sigma^\nu  \bar{\sigma}^\mu\: =\:
 2 \eta^{\mu\nu} {\bf 1}_2\,,\qquad
 \bar{\sigma}^\mu \sigma^\nu  \, +\, \bar{\sigma}^\nu \sigma^\mu  \: =\:
 2 \eta^{\mu\nu} {\bf 1}_2\,,
\end{equation}
with $\eta^{\mu\nu} = \text{diag}(1,-1,-1,-1)$. Moroever, the following completeness relation involving the two set of generators lies at the heart of the bilinear formalism:
\begin{equation}
  \label{eq:Complete}
 (\bar{\sigma}_\mu)_{ij}\, (\sigma^\mu)_{kl} \ =\  
 (\sigma_\mu)_{ij}\, (\bar{\sigma}^\mu)_{kl}\ =\ 2\, \delta_{il}\,\delta_{kj}\, ,
\end{equation}
with $i,j,k,l =1,2$.
In addition, the two set of generators, $\bar{\sigma}^\mu$ and $\sigma^\mu$, are related to one another through,
\begin{equation}
    \label{eq:s2sbars2}
   \sigma^2\, \bar{\sigma}^\mu\, \sigma^2\: =\: (\sigma^\mu)^{\sf T}\,. 
\end{equation}
As we will see below, the two spinorial algebras will form
the basis for the Dirac algebra that we will utilise in the bilinear field space. 

Given the scalar multiplet ${\bf \Phi}$ as defined
in~\eqref{eq:Phi}, it is straightforward to go over to the bilinear field space which   realizes  an   O(1,3)   symmetry  group~\cite{Ivanov:2007de}. To this end, we  define the field-space four vector:
\begin{equation}
  \label{eq:RA}
  R^\mu\: \equiv\:
  {\bf \bar{\Phi}}\, \Gamma^\mu\, {\bf \Phi}\ =\ 
2\,\left( \begin {array}{c} \phi_1^{\dagger}
 \phi_1+\phi_2^{\dagger} \phi_2\\\noalign{\medskip}\phi_1^{\dagger}
 \phi_2+\phi_2^{\dagger}
 \phi_1\\\noalign{\medskip}-i\left[\phi_1^{\dagger}
 \phi_2-\phi_2^{\dagger}
 \phi_1\right]\\\noalign{\medskip}\phi_1^{\dagger}
 \phi_1-\phi_2^{\dagger} \phi_2
\end {array} \right) \; ,
\end{equation}
with $\mu = 0,1,2,3$. We use letters from the middle of the 
Greek alphabet, like $\mu\,,\nu$, to refer to the components of the bilinear field-space four-vector, e.g.~$R^\mu$, as well as to distinguish them from spacetime coordinates, like $x^{\alpha}$, for which we use letters from the beginning of the alphabet. The  $8\times 8$-dimensional matrices $\Gamma^\mu$  may be expressed as follows:
\begin{equation}
   \label{eq:Gammamu} 
   \Gamma^\mu \: \equiv\: 
   \bar{\gamma}^\mu\otimes {\bf 1}_2\: =\: \left(
     \begin{array}{cc}
       {\bf 0}_2 & \bar{\sigma}^\mu\\
       \sigma^\mu & {\bf 0}_2
      \end{array}\right) \otimes {\bf 1}_2\;, 
\end{equation}
which obey the standard Dirac algebra,
\begin{equation}
  \label{eq:Dirac}
  \{ \Gamma^\mu\,,\, \Gamma^\nu\}\: =\: \Gamma^\mu\, \Gamma^\nu\: +\:
  \Gamma^\nu\, \Gamma^\mu\: =\: 2\,\eta^{\mu\nu}\, {\bf 1}_8\;.
\end{equation}  
Apart from a trivial swap between $\sigma^\mu$ and $\bar{\sigma}^\mu$ in~\eqref{eq:Gammamu}, $\{\bar{\gamma}^\mu\} = 
\{\gamma_\mu\}$ are identical to the standard Dirac matrices in the Weyl representation. Correspondingly, the Lorentz-dual representation~${\bf \bar{\Phi}}$ of the scalar multiplet~${\bf \Phi}$ is defined as
\begin{equation}
 \label{eq:Phibar}
 {\bf \bar{\Phi}}\ \equiv\ {\bf \Phi}^\dagger\, \Gamma^0\ =\
 \Big(\!\!\begin {array}{ccc}
 \big(\!\!\begin{array}{cc}
\phi^{\sf T}_1 i\sigma^2 & \phi^{\sf T}_2 i\sigma^2 
 \end{array}\!\!\big) \big(i\sigma^2\otimes {\bf 1}_2\big) &
 \phi^\dagger_1 & \phi^\dagger_2 \end{array}
 \!\!\Big)\,,
 \end{equation}
 in close analogy to the standard formulation for Dirac or Majorana
 fermions. We should stress here that the equivalence between this Dirac-algebra bilinear formalism and the one given previously in~\cite{Battye:2011jj,Pilaftsis:2011ed}  can be explicitly traced to the key relation~\eqref{eq:s2sbars2}.
 
 In the present Dirac formulation, a general U(1)$_Y$-invariant  SL($2\,,\mathbb{C}$)-transformation of the scalar multiplet  ${\bf \Phi}$ can be performed, in terms of the matrices $\Sigma^{\mu\nu} \equiv \frac{i}{2}\, [\Gamma^\mu\,, \Gamma^\nu]$, as 
 \begin{equation}
   \label{eq;Phiprime}
 {\bf \Phi}'\ =\ \exp \bigg(\!\!-\frac{i}{4}\, \Sigma^{\mu\nu}\,  \omega_{\mu\nu}\bigg)\: {\bf \Phi}\,,
\end{equation}
where $\omega_{\mu\nu} = -\omega_{\nu\mu}$ are the six group parameters of  the SL($2\,,\mathbb{C}$) group~\cite{Bjorken:100769}. As a consequence of such a transformation, we recover the known result~\cite{Ivanov:2006yq}:
 $R'^\mu \equiv {\bf \bar{\Phi}'}\, \Gamma^\mu\, {\bf \Phi'} =
 \Lambda^\mu_{\ \nu}\, R^\nu$, with
 $\Lambda^\mu_{\ \nu} \in \text{O}(1,3)$, i.e.~$R^\mu$ is a proper
 four-vector~\footnote[1]{Strictly speaking, the maximal U(1)$_Y$-invariant field reparameterisation group in the scalar-doublet field-space $\phi_{1,2}$ is GL($2\,,\mathbb{C}$) and includes an overall scaling factor~$e^{\sigma}$ (with $\sigma\in \mathbb{R}$), such that $R'^\mu = e^{2\sigma}\,
 \Lambda^\mu_{\ \nu}\, R^\nu$~\cite{Battye:2011jj}.}. 
 
Following~\cite{Battye:2011jj,Pilaftsis:2011ed}, we may analogously define an operation of charge conjugation that  acts on the 8D scalar multiplet~${\bf \Phi}$,  
 \begin{equation}
  \label{eq:MajoranaPhi}
{\bf \Phi}^C\ \equiv\ C\,  {\bf \bar{\Phi}}^{\sf T}\ =\ {\bf \Phi} \,, 
\end{equation}
where $C = -i\Gamma^2\Gamma^0 C_S$, with $C_S = \bar{\gamma}^5\otimes (i\sigma^2)$ and $\bar{\gamma}^5 = -i\bar{\gamma}^0 \bar{\gamma}^1\bar{\gamma}^2\bar{\gamma}^3$. Evidently, this operation exemplifies
the Majorana structure of the 8D scalar multiplet ${\bf \Phi}$. 
Like Majorana fermions and up to a relative minus sign due to the difference between Fermi and Bose statistics, one can show that the 2HDM-scalar multiplets ${\bf \Phi} (x_{1,2})$ at two different spacetime locations $x_{1,2}$ obey the useful symmetric property:
\begin{equation}
    \label{eq:MajPhi}
 {\bf \bar{\Phi}}(x_1)\, \Gamma^\mu\, {\bf \Phi}(x_2)\ =\   
 {\bf \bar{\Phi}}^C(x_1)\, \Gamma^\mu\, {\bf \Phi}^C(x_2)\ =\ 
 {\bf \bar{\Phi}}(x_2)\, \Gamma^\mu\, {\bf \Phi}(x_1)\,.
\end{equation}
In the above, we made use of the identities: $C^{-1}\! =\! C^{\sf T}\! =\! C\! =\! \Gamma^0C^\dagger\Gamma^0\! \equiv\! \bar{C}$ and $C^{-1}\Gamma^\mu C = (\Gamma^\mu )^{\sf T}$.

We are now in a position to rewrite the 2HDM Lagrangian
in~\eqref{eq:L2HDM} using the Dirac-algebra formalism that we have
established here. To start with,  we employ the four-vector $R^\mu$, in order to rewrite the scalar potential $V$ in \eqref{eq:V2HDM}  in the equally familiar quadratic form~\cite{Maniatis:2006fs,Nishi:2006tg,Ivanov:2006yq},
\begin{equation}
  \label{eq:VR}
V\ =\ \frac{1}{2}\,M_\mu\, R^\mu\: +\: \frac{1}{4}\,L_{\mu\nu}\,
R^\mu R^\nu\; ,
\end{equation}
where 
\begin{eqnarray}
M_\mu \!\!&=&\!\!  \frac{1}{2}\,\left(\! \begin {array}{cccc} m_{11}^2 +
 m_{22}^2 & \:2\mathrm{Re}(m_{12}^2) &-2\mathrm{Im}(m_{12}^2) &m_{22}^2 -
 m_{11}^2\end {array}\!\right)\; , \\[3mm]
L_{\mu\nu} \!\!&=&\!\! L_{\nu\mu}\ =\ \frac{1}{4}\,\left(\! \begin {array}{cccc} \lambda_1 +
 \lambda_2 + \lambda_3&\mathrm{Re}(\lambda_6 +
 \lambda_7)&-\mathrm{Im}(\lambda_6 + \lambda_7)&\lambda_1 - \lambda_2\\
\noalign{\medskip}\mathrm{Re}(\lambda_6 +
 \lambda_7)&\lambda_4+\mathrm{Re}(\lambda_5)&
 -\mathrm{Im}(\lambda_5)&\mathrm{Re}(\lambda_6 
 - \lambda_7)\\
\noalign{\medskip}-\mathrm{Im}(\lambda_6 +
\lambda_7)&-\mathrm{Im}(\lambda_5)&\lambda_4 -
\mathrm{Re}(\lambda_5)&-\mathrm{Im}(\lambda_6 -
\lambda_7)\\
\noalign{\medskip}\lambda_1 - \lambda_2&\mathrm{Re}(\lambda_6 -
\lambda_7)&-\mathrm{Im}(\lambda_6 - 
            \lambda_7)&\lambda_1+\lambda_2-\lambda_3
\end{array} \!\right)\,.\qquad
\end{eqnarray}

We now turn our attention to the gauge-kinetic terms of the scalar doublets. In the Dirac-algebra formalism under study, these can be expressed as follows:
\begin{equation}
  \label{eq:Lkin}
{\cal L}^{\rm S}_{\rm kin}\ =\ (D_\alpha\phi_1)^\dagger (D^\alpha \phi_1)\, +\,
  (D_\alpha\phi_2)^\dagger (D^\alpha \phi_2)\ =\
  \frac{1}{2}\,(\widehat{D}_\alpha {\bf \Phi})^\dagger 
(\widehat{D}^\alpha{\bf \Phi})\ =\ -\,\frac{1}{2}\,\zeta_\mu \, {\bf \bar{\Phi}}\,\Gamma^\mu
\widehat{\Box}\, {\bf \Phi}\,,
\end{equation}
where $\widehat{D}_\alpha = {\bf 1}_8\,\partial_\alpha + \frac{i}{2} g\, W^i_\alpha\,  ({\bf
  1}_4\otimes \sigma^i) + \frac{i}{2}g'\,B_\alpha\,( \sigma^3\otimes {\bf 1}_4)$ is the covariant spacetime derivative in the 8D ${\bf
  \Phi}$-space with respect to the SM gauge group,
and $\widehat{\Box} \equiv \widehat{D}_\alpha\widehat{D}^\alpha$ is
the corresponding SM-gauge covariant Laplacian. To obtain the last
equality in~\eqref{eq:Lkin}, we use the identity: $\partial_\alpha(
{\bf \Phi}^\dagger \widehat{D}^\alpha{\bf \Phi})\, =\,
(\widehat{D}_\alpha {\bf \Phi})^\dagger 
(\widehat{D}^\alpha{\bf \Phi}) + {\bf \Phi}^\dagger(\widehat{\Box}\,
{\bf \Phi})$, to drop a total spacetime derivative. In addition, following~\cite{Ivanov:2007de}, we
have introduced the wave-function four-vector $\zeta^\mu$,
which reads:
\begin{equation}
   \label{eq:Zmu}
  \zeta^\mu\ =\ (1\,, 0\,,0\,, 0)
\end{equation}
at the tree level in the canonical field basis. However, beyond the Born approximation, the components of $\zeta^\mu$ can change drastically~\cite{Bednyakov:2018cmx}, due to wave-function renormalisation of the bare scalar doublets, $\phi_{1,2}$, in terms of their renormalised counterparts, $\phi^{\rm R}_{1,2}$. Following the standard procedure, we have:~$\phi_i = (Z^{1/2}_\phi)_{ij}\,\phi^{\rm R}_j$, where $(Z^{1/2}_\phi)_{ij}$ are the usual wave-function renormalisations (with~$i,j =1,2$) and summation over the repeated index~$j$ is implied. Then, the $2\times 2$ wave-function renormalisation matrix $Z^{1/2}_\phi$ can be related to $\zeta^\mu$ through:
\begin{equation}
   \label{eq:Zphi-mu}
 Z^{1/2\,\dagger}_\phi\, Z^{1/2}_\phi \: =\: \zeta_\mu\,\sigma^\mu\, .
\end{equation}
The components of $\zeta^\mu$ will start receiving UV-infinite contributions from fermions and gauge bosons at one-loop level, and from scalars at two loops. To~deal with the UV infinities, mixing renormalisation between the scalar doublets must also be considered~\cite{Pilaftsis:1997dr}. Furthermore, given that $R^0>0$, the absence of ghosts associated with negative kinetic terms will impose the two `time'-like constraints: (i)~${\zeta^0> 0}$  and (ii)~$\zeta^2 > 0$. As~we will see in the next section, the inclusion of the wave-function four-vector, $\zeta^\mu$, is essential to obtain an O(1,3)-covariant expression for the effective potential.

Let us finally discuss the last gauge-invariant term of the 2HDM Lagrangian~\eqref{eq:L2HDM}, i.e.~the Yukawa Lagrangian~${\cal L_{\rm Y}}$. Considering only quark states for simplicity, this last term may conveniently\- be written as~\cite{BhupalDev:2014bir}
\begin{equation}
  \label{eq:LYuk}
- {\cal L}_{\rm Y}\ =\ \overline{Q}_L\, {\cal M}_Q[{\bf \Phi}]\: Q_R\ +\ {\rm H.c.}\,, 
\end{equation}
where $Q_{L\,(R)} = ( u_{L\,(R)}\,, d_{L\,(R)})^{\sf T}$ and
\begin{equation}
   \label{eq:calMQ}
{\cal M}_Q[{\bf \Phi}]\ =\ \left(\! \begin{array}{cc}
 h^u_i i\sigma^2 \phi^*_i    &  h^d_i \phi_i
\end{array}
\!\right)  
\end{equation}
is an SU(2)$_L$ gauge-covariant, ${\bf \Phi}$-dependent quark-mass matrix, while summation over the repeated index $i=1,2$ is understood. In~writing ${\cal L_{\rm Y}}$ in~\eqref{eq:LYuk}, we suppressed the inter-generational indices of the down- and up-quark Yukawa-coupling matrices,~$h^d_{1,2}$ and $h^u_{1,2}$, associated with the Higgs doublets $\phi_{1,2}$ and their hypercharge-conjugate counterparts,~$i\sigma^2 \phi^*_{1,2}$.  Although the Yukawa Lagrangian~${\cal L_{\rm Y}}$ being linear in $\phi_{1,2}$ cannot be written in bilinear form, we will still be able to obtain an O(1,3)-invariant result for
the effective potential from fermion loops, thanks to the covariant SL(2,$\mathbb{C}$)-representation of ${\cal M}_Q[{\bf \Phi}]$ in~\eqref{eq:calMQ}. As we will see in Section~\ref{subsec:F},  this additional feature will prove crucial in arriving at a covariant effective potential after including fermion loops.

\section{Covariant One-Loop Effective Potential}\label{sec:effpot}

We will now apply the Dirac algebra formalism in order to obtain a fully $\text{O}(1,3)$-covariant effective potential at the one-loop level~\cite{Coleman:1973jx}. To this end, we employ the general functional formula~\cite{Jackiw:1974cv,Ramond:1997pw,Zinn-Justin:2002ecy}:
\begin{equation} 
	\label{eq:V1loop}
V_{\mathrm{eff}}\ =\  -\,C_{\rm s}\, \frac{i \hbar}{2}\: 
\ln \bigg(\frac{\det H_{\varphi_{1} \varphi_{2}}(\varphi)}{\det H_{\varphi_{1} \varphi_{2}}(0)} \bigg) 
\ =\  -\,C_{\rm s}\, \frac{i \hbar}{2}\:  \Big(
\mathrm{Tr} \ln H_{\varphi_{1} \varphi_{2}}(\varphi) -
\mathrm{Tr} \ln H_{\varphi_{1} \varphi_{2}}(0) \Big)\; ,   
\end{equation}
where $C_{\rm s}  = +1~(-1)$  for the generic fields $\varphi_{1,2}$   obeying   the   Bose--Einstein
(Fermi--Dirac)   statistics, and 
$H_{\varphi_{1} \varphi_{2}} (\varphi )$  is the second derivative  of the
classical action~${S = \int d^4 x\, {\cal L}}$, 
\begin{equation}
	\label{eqn:H}
H_{\varphi_{1} \varphi_{2}}(\varphi )\:  = \:  \frac{\delta^{2}
  S}{\delta  \varphi_{1}   (x_1)\delta  \varphi_{2} (x_{2})} \; , 
\end{equation}
which is the {\em Hessian} of the action and also termed the {\em inverse background-field propagator}.
In~\eqref{eqn:H},  $\varphi_{1,2}$ collectively denotes each of  the fields,
\begin{displaymath}
\{ {\bf \Phi},  W^{i}_{\mu}, B_{\mu}, u_{i}, d_{i}, e_{i}, \nu_{i} \}\,, 
\end{displaymath}
as well as the Goldstone bosons and the electroweak ghosts which do not contribute to $V_{\text{eff}}$ in the Landau gauge. In~\eqref{eq:V1loop}, 
$\varphi$ are the background fields which are taken to be the elements of the 8D ${\bf \Phi}$-multiplet. Moreover, the symbol trace ($\mathrm{Tr}$) in \eqref{eq:V1loop} acts over the configuration space and all internal degrees of freedom. 

To evaluate the one-loop effective potential in~\eqref{eq:V1loop}, we will frequently  make use of a more convenient representation given by Eq.~(B.4) in~\cite{Alexander-Nunneley:2010tyr},  
\begin{equation} 
	\label{eq:V1loopeff}
V_{\mathrm{eff}}(\varphi)\: =\: - C_{\rm s}\, \frac{i}{2}  \int^{1}_{0}
dx\, \mathrm{Tr} \bigg[ \frac{H_{\varphi_{1} \varphi_{2}}(\varphi ) -
    H_{\varphi_{1} \varphi_{2}}(0)}{x \left( H_{\varphi_{1}
      \varphi_{2}}(\varphi) -  H_{\varphi_{1} \varphi_{2}}(0)
    \right) + H_{\varphi_{1} \varphi_{2}}(0)}  \bigg] \; . 
\end{equation} 
Note that the validity of this representation relies on the condition that the commutator between the two Hessians $H_{\varphi_{1} \varphi_{2}}(\varphi )$ and $H_{\varphi_{1} \varphi_{2}}(0)$ vanishes, i.e.
\begin{equation}
   \label{eq:HessCom}
[H_{\varphi_{1} \varphi_{2}}(\varphi )\,, H_{\varphi_{1} \varphi_{2}}(0)]\ =\ 0\,.
\end{equation}
Since $H_{\varphi_{1} \varphi_{2}}(0)$ often becomes proportional to
the identity operator, this condition is applicable for most cases of
interest to us. An exception  is the scalar-boson
Hessian~\eqref{eq:HessPhi} to be discussed in the next subsection,
where explicit use of the Dirac algebra formalism is made.
In the momentum $k$-space  of $d = 4  - 2 \varepsilon$  dimensions,  the expression in~\eqref{eq:V1loopeff} becomes 
\begin{equation} 
	\label{eq:V1loopk}
V_{\mathrm{eff}}(\varphi)\: =\: -\,C_{\rm s}\, \frac{i}{2} \int^{1}_{0}
dx\, \int \frac{\mu^{2\varepsilon}\,d^d k}{(2\pi)^d}\:\mathrm{tr} \bigg[ \frac{H_{\varphi_{1}
      \varphi_{2}}(\varphi) -  H_{\varphi_{1} \varphi_{2}}(0)}{x
    \left( H_{\varphi_{1} \varphi_{2}}(\varphi) -  H_{\varphi_{1}
      \varphi_{2}}(0) \right) + H_{\varphi_{1} \varphi_{2}}(0)}
  \bigg]\,, 
\end{equation} 
where $\mu$ is the so-called 't Hooft mass in the Dimensional Regularisation (DR) scheme~\cite{tHooft:1973mfk}, and the operation denoted by 
`$\mathrm{tr}$' stands for the  trace over  the internal degrees of  freedom only, such as the polarisations of  the gauge bosons, the spinor components of the fermions or the Yukawa coupling matrices. 

The one-loop effective  potential of the 2HDM can  now be calculated by applying \eqref{eq:V1loopk} to the scalars (S), gauge bosons (GB),  and fermions (F) individually, i.e.
\begin{equation}
 	\label{eq:Vefftotal}
 V_{\mathrm{eff}}({\bf \Phi})  \: =\:
 V^{\rm S}_{\mathrm{eff}}({\bf \Phi})\, +\,
 V^{\rm GB}_{\mathrm{eff}}({\bf \Phi})\, 
 +\, V^{\rm F}_{\mathrm{eff}}({\bf \Phi})\; ,    
\end{equation}
where the Majorana-scalar multiplet~${\bf \Phi}$ replaces the generic background field $\varphi$. The aim of this exercise is to re-express the one-loop effective potential in terms of $R^\mu$ only, in an O(1,3)-invariant manner, i.e.~$V_{\mathrm{eff}}({\bf \Phi})  = V_{\mathrm{eff}}[R({\bf \Phi})]$.

\subsection{Scalar Loops}\label{subsec:S}

Let us first consider the scalar-doublet sector only. In the ${\bf \Phi}$-space, the  Hessian $H^{\rm S}_{{\bf \bar{\Phi}{\bf \Phi}}}({\bf \Phi})$ takes on the 
$\text{O}(1,3)$-covariant form in the momentum representation, 
\begin{align}
 \label{eq:HessPhi}
H^{\rm S}_{{\bf \bar{\Phi}{\bf \Phi}}}({\bf \Phi}) =&\ A_\mu [R]\, \Gamma^\mu\: -\: {\bf B}\,,
\end{align}
with the identifications: 
\begin{equation}
   \label{eq:DefAB}
A_\mu [R]\: \equiv\: \zeta_\mu k^2\, -\, M_\mu\, -\, L_{\mu\nu}\, R^\nu\, ,\qquad {\bf B}\: \equiv\: 2\,(\Gamma^\mu {\bf \Phi})\, L_{\mu\nu}\, ({\bf \bar{\Phi}} \Gamma^\nu)\,.
\end{equation}
In the above, $A_\mu [R]$ is a four-vector depending on both the four-momentum $k$ and four-vector~$R^\mu$, and ${\bf B}$ is an $8\times 8$-dimensional matrix in the ${\bf \Phi}$-space.
We note that the decomposition in~\eqref{eq:HessPhi} agrees with~\cite{Cao:2022rgh} which was obtained by adopting the Clifford-algebra formulation of~\cite{Degee:2009vp}. With the help of the Dirac algebra~\eqref{eq:Dirac}, we can readily invert the matrix: $\slashed{A}[R] \equiv A_\mu[R]\, \Gamma^\mu$, by following the old textbook recipe~\cite{Bjorken:100769}, i.e.~$\slashed{A}{}^{-1} = \slashed{A}/\slashed{A}{}^2 = \slashed{A}/A^2$, with $A^2 \equiv A_\mu A^\mu$. 

To bosonise the leading momentum-dependent fermionic part $\slashed{A}[R]$ of the Hessian $H^{\rm S}_{{\bf \bar{\Phi}{\bf \Phi}}}$, we rewrite the logarithm of its functional determinant,
$\ln \det (H^{\rm S}_{{\bf \bar{\Phi}{\bf \Phi}}})$, as follows~\footnote[2]{Another way to bosonise the leading momentum-dependent term will be to multiply $\det \big(\slashed{A} - {\bf B})$ by a field-independent and O(1,3)-invariant constant, e.g.~by~$\det\slashed{\zeta} = (\zeta^2)^{{\rm Tr}{\bf 1}/2}$. Here we will not follow this route.}:
\begin{eqnarray}
  \label{eq:LnDetHS}
\ln \det \big(\slashed{A} - {\bf B}) \!&=&\!
\ln\det\slashed{A}\: +\: \ln\det \big({\bf 1} - \slashed{A}{}^{-1}{\bf B} \big)\ =\  \frac{1}{2} \ln\det \slashed{A}{}^2\: +\: 
\text{Tr} \ln \big({\bf 1} - \slashed{A}{}^{-1}{\bf B} \big)\nonumber\\
\!&=&\! \frac{1}{2}\, \text{Tr} \ln  (A^2{\bf 1})\  -\ 
\sum_{n=1}^\infty \frac{1}{n}\:\text{Tr} \bigg(\frac{(\slashed{A}\, {\bf B})^n}{(A^2)^n}\bigg)\, . 
\end{eqnarray}
Here, we remind the reader that all traces (Tr) are to be taken with respect to the configuration space, including internal degrees of freedom. In this regard, we proceed in good faith by assuming the absence of the so-called {\em multiplicative anomalies} for non-{\em trace class} operators~(see~\cite{Evans:1998pd,Niedermaier:2010zz} and references therein). These anomalies refer to the violation of the properties of determinants for two finite matrices $A_1$ and $A_2$, e.g.~$\det (A_1 A_2) = \det A_1 \det A_2$, which should also hold true for the {\em infinite} dimensional matrices, like $\slashed{A}$ and ${\bf B}$, as these are defined on an infinite configuration space. Nevertheless, as argued in~\cite{Evans:1998pd} by analysing simple examples, such anomalies bear no physical significance, as they only seem to affect the scheme of renormalisation or the choice of the 't Hooft mass $\mu$ in~\eqref{eq:V1loopk}. 

As can be seen from the last equality in~\eqref{eq:LnDetHS}, the 
scalar-doublet contribution to the effective potential, $V^{\rm S}_{\mathrm{eff}}({\bf \Phi})$, consists of two terms. The first term has a simple spinorial form proportional to ${\bf 1}_8$ and can be computed by means of~\eqref{eq:V1loopk}, i.e.
\begin{equation}
   \label{eq:TrLnA2}
 \frac{1}{2}\, \text{Tr} \ln  (A^2{\bf 1})\ =\  \frac{\text{tr} {\bf 1}_8}{2}\,
 \int^{1}_{0}
dx\, \int \frac{\mu^{2\varepsilon}\, d^d k}{(2\pi)^d}\: \frac{\delta A^2 [R]}{A^2[0]\,+\, x\,\delta A^2 [R]}\ , 
\end{equation}
where $(\text{tr} {\bf 1}_8)/2 = 4$, and 
$A^2[R]\equiv A[R]\!\cdot\! A[R]$ and $\delta A^2 [R] \equiv A^2[R] - A^2[0]$. Hereafter, a dot~($\cdot$) indicates an $\text{O}(1,3)$-invariant contraction with respect to the available Lorentz indices in the bilinear $R$-space. Consequently, we have 
\begin{eqnarray}
    \label{eq:dA2}
A^2[0] &=& \zeta^2\, (k^2)^2\: -\: 2\,(\zeta\!\cdot\!M)\, k^2\: +\: M^2\,,\nonumber\\
\delta A^2 [R] & = & -\, 2\,(\zeta\!\cdot\!L\!\cdot\!R)\,k^2 \: +\:
2\,(M\!\cdot\!L\!\cdot\!R)\: +\: (R\!\cdot\!L^2\!\cdot\!R)\,.
\end{eqnarray}
Some remarks are now in order. First, we observe that the inclusion of the wave-function vector $\zeta^\mu$ [cf.~\eqref{eq:Zmu}] plays an instrumental role in arriving at a fully $\text{O}(1,3)$-invariant expression in~\eqref{eq:TrLnA2} in the bilinear $R$-space. Second, we notice that unlike~\cite{Cao:2023kgq}, \eqref{eq:TrLnA2} does not suffer from IR divergences, since  
both $A^2[0]$ and $\delta A^2 [R]$ do not vanish in general 
as the loop momentum goes to zero, i.e.~$k\to 0$. Third, as we will see below, our improved Dirac-algebra approach leads to a different and less convoluted re-organisation of the contributing $R$-dependent terms. Finally, an explicit analytic computation of~\eqref{eq:TrLnA2} is straightforward and can be done using partial fraction decomposition, but it goes beyond the scope of the present article and could be carried out elsewhere.

We now turn our attention to the second term of the last equality in~\eqref{eq:LnDetHS}. To gain some insight of its analytical structure, we start by considering the first term $n=1$ of the infinite sum over $n$. By considering only internal degrees of freedom, the $n=1$ term can be calculated~as 
\begin{equation}
    \label{eq:n1term}
 {\rm tr}\, \slashed{A}\,{\bf B}\ =\ 2\,A_\rho\,
 \big(\bar{\bf \Phi}\,\Gamma^\nu\Gamma^\rho\Gamma^\mu\,
 {\bf \Phi}\big)\, L_{\mu\nu}\ =\
 4\, (A\!\cdot\!L\!\cdot\!R)\: -\: 2\,(A\!\cdot\!R)\,{\rm tr}\,L\;  ,
\end{equation}
where we employed \eqref{eq:RA} to go from the ${\bf \Phi}$- to the bilinear $R$-space after the following identity for a product of three Dirac matrices was utilised:
\begin{equation}
  \label{eq:DiracId}
 \Gamma^\nu\Gamma^\rho\Gamma^\mu\ =\ \eta^{\nu\rho}\,\Gamma^\mu\: +\: \eta^{\rho\mu}\,\Gamma^\nu\: -\: \eta^{\nu\mu}\,\Gamma^\rho\: +\: i\, \varepsilon^{\nu\rho\mu\sigma}\Gamma_\sigma \Gamma^5\,, 
\end{equation}
with the conventions: $\Gamma^5 = i\Gamma^0\Gamma^1\Gamma^2\Gamma^3 = \bar{\gamma}^5\otimes {\bf 1}_2$ and $\varepsilon^{0123} = +1$ for the 4D Levi-Civita symbol. Moreover, we employed the property: ${\bf\bar{\Phi}}\,\Gamma^\mu\Gamma^5\,{\bf \Phi} = 0$, along with the abbreviations: ${{\rm tr}\,L = \eta^{\mu\nu}L_{\nu\mu} = L^\mu_{\ \mu}}$, ${{\rm tr}\,L^2 = L^\mu_{\ \nu}}\, L^\nu_{\ \mu}$, 
${{\rm tr}\,L^3 = L^\mu_{\ \nu_1}} L^{\nu_1}_{\ \nu_2} L^{\nu_2}_{\ \mu}$ etc.
After some familiarisation,  the $n^{\rm th}$-order summand may therefore be computed as follows:
\begin{eqnarray}
    \label{eq:nallterm}
{\rm tr}\, (\slashed{A}\,{\bf B})^n\!&=&\! 
\big(\bar{\bf \Phi}\,\Gamma^{\nu_n}\slashed{A}\Gamma^{\mu_1}\,{\bf \Phi}\big)\, L_{\mu_1\nu_1}\, 
\big(\bar{\bf \Phi}\,\Gamma^{\nu_1}\slashed{A}\Gamma^{\mu_2}\,{\bf \Phi}\big)\, L_{\mu_2\nu_2}\, 
\cdots
\big(\bar{\bf \Phi}\,\Gamma^{\nu_{n-1}}\slashed{A}\Gamma^{\mu_n}\,{\bf \Phi}\big)\, L_{\mu_n\nu_n}\, \nonumber\\
\!&=&\!\ 2^n \prod_{i=1}^n \Big(A^{\nu_i} (R\!\cdot\!L)_{\nu_{i+1}}
+ R^{\nu_i} (A\!\cdot\!L)_{\nu_{i+1}} - (A\!\cdot\!R)\, L^{\nu_i}_{\ \nu_{i+1}}\Big)\: \equiv\: F_n[R]\,,
\end{eqnarray}
with the index identification: $\nu_{n +1} \equiv \nu_1$.
Notice that only specific contractions among the tensors, 
$R_\mu$, $A_\mu[R]$ and $L_{\mu\nu}$,  are allowed in the reparameterisation-invariant expression $F_n[R]$ in~\eqref{eq:nallterm}.
Taking~\eqref{eq:LnDetHS}, \eqref{eq:TrLnA2}  and~\eqref{eq:nallterm} into account, the manifestly O(1,3)-invariant scalar contribution to the 2HDM one-loop effective potential takes on the form:
\begin{eqnarray}
  \label{eq:VeffScalar}
V^{\rm S}_{\mathrm{eff}}[R] \ =\
\int^{1}_{0}\!dx 
\int \frac{\mu^{2\varepsilon}\, d^d k}{(2\pi)^d\,i}\: \bigg(\, \frac{2\,\delta A^2}{A^2_0\,+\, x\,\delta A^2}\ -\
\sum_{n=1}^\infty \frac{1}{2n}\,\frac{F_n[R]}{(A^2)^n}\,\bigg)\,,\qquad 
\end{eqnarray}
where $A_\mu = A_\mu[R]$, $A^2_0 \equiv A^2[0]$ and 
$\delta A^2 = \delta A^2[R]$, according to our definitions  in~\eqref{eq:DefAB} and~\eqref{eq:dA2}.  It~is~important to reiterate here that unlike~\cite{Cao:2022rgh,Cao:2023kgq}, $V^{\rm S}_{\mathrm{eff}}[R]$, as stated in~\eqref{eq:VeffScalar}, does not exhibit IR poles, since the expressions~$A^2_0$ and~$A^2$ that appear in the denominators within the integrand do not vanish in general as the loop momentum $k$ goes to zero. 

Let us now evaluate the UV behaviour of~$V^{\rm S}_{\mathrm{eff}}({\bf \Phi})$ given in~\eqref{eq:VeffScalar}.  To this end, we first observe from~\eqref{eq:dA2} a quartic loop-momentum dependence for $A^2_0 \propto (k^2)^2$, whilst for~$A_\mu \propto k^2$ and $\delta A^2 \propto k^2$, the $k$-dependence is only quadratic in the UV limit of the theory, as $|k| \to \infty$. Consequently, the infinite sum in the integrand on the RHS of~\eqref{eq:VeffScalar}
becomes UV finite for all terms~$n> 2$. 
In the DR scheme with $d=4-2\varepsilon$ dimensions, the loop integrals of interest are:
\begin{eqnarray}
    \label{eq:Iloop1}
I_1 \!&\equiv&\! \int \frac{\mu^{2\varepsilon}\, d^d k}{(2\pi)^d\,i}\, \frac{1}{k^2 - \Delta^2 +i\epsilon}\ =\ \frac{\Delta^2}{16\pi^2}\,\bigg[\, \frac{1}{\varepsilon}\, +\, 1\: -\:\ln \bigg(\frac{\Delta^2}{\bar{\mu}^2}\bigg)\bigg]\,,\\
    \label{eq:Iloop2}
 I_2 \!&\equiv&\! \int \frac{\mu^{2\varepsilon}\, d^d k}{(2\pi)^d\,i}\, \frac{1}{(k^2 - \Delta^2 +i\epsilon)^2}\ =\ \frac{1}{16\pi^2}\, \bigg[\, \frac{1}{\varepsilon}\: -\:\ln \bigg(\frac{\Delta^2}{\bar{\mu}^2}\bigg)\bigg]\,,   
\end{eqnarray}
where $\Delta^2$ is an arbitrary squared mass parameter, and $\ln\bar{\mu}^2 = -\gamma + \ln (4\pi\mu^2)$, with $\gamma \approx 0.5772$ being the Euler-Mascheroni constant. With the help of the loop integrals~$I_{1,2}$, we may compute the UV-divergent part of~$V^{\rm S}_{\mathrm{eff}}({\bf \Phi})$,
\begin{eqnarray}
    \label{eq:VSeffUV}
V^{\rm S\,, UV}_{\mathrm{eff}}[R]
\!\!&=&\!\! -\,\frac{1}{16\pi^2\,\varepsilon}\, 
\Big[\, 12\, (\zeta\!\cdot\! M)\,(\zeta\!\cdot\!L\cdot\!R)
-\: 2\,(\zeta\!\cdot\!M)\, (\zeta\!\cdot\!R)\:\widehat{L}\:
-\: 6\, \zeta^2(M\!\cdot\!L\cdot\!R)\: +\: \zeta^2(M\!\cdot\!R)\:\widehat{L}\nonumber\\
\!&&\!
+\: 10\,(\zeta\!\cdot\!L\cdot\!R)^2\: -\: 4\, (\zeta\!\cdot\!R)\,(\zeta\!\cdot\!L^2\cdot\!R)\: +\: 
2\, (\zeta\!\cdot\!L\cdot\!\zeta)\,(R\!\cdot\!L\cdot\!R)\\
\!&&\!-\: 2\, (\zeta\!\cdot\!R)\, (\zeta\!\cdot\!L\cdot\!R)\:\widehat{L}\: +\: 
(\zeta\!\cdot\!R)^2\:\widehat{L}^2\: -\: 4\,\zeta^2 (R\!\cdot\!L^2\!\cdot\!R)\: +\: \zeta^2 (R\!\cdot\!L\!\cdot\!R)\:\widehat{L}\, \Big]\,,\qquad\nonumber
\end{eqnarray}
with the short-hand: $\widehat{L}^k \equiv {\rm tr}L^k$ ($k=1,2$). As expected, \eqref{eq:VSeffUV} is manifestly O(1,3)-invariant. 

Knowing the explicit form of $V^{\rm S\,, UV}_{\rm eff}[R]$ as stated in~\eqref{eq:VSeffUV}, it is not difficult to derive the O(1,3)-covariant $\beta$- and $\gamma$-functions 
governing the RG-running of the tensors~$L_{\mu\nu}$ and~$M_\mu$. Given that the wave-function four-vector~$\zeta^\mu$ does not renormalise from pure scalar loops, i.e.~$d\zeta^\mu/dt = 0$ (with $t\equiv \ln\mu^2$), we have 
\begin{eqnarray}
    \label{eq:betaL}
(\beta_L)_{\mu\nu} \!&\equiv&\! \frac{dL_{\mu\nu}}{dt}\ =\
-\:\lim_{\varepsilon \to 0}\,2\varepsilon\: \frac{\partial^2V^{\rm S,UV}_{\rm eff}[R]}{\partial R^\mu\partial R^\nu}\ =\ \frac{1}{16\pi^2}\,\Big[\, 40\,(\zeta\!\cdot\!L)_\mu (\zeta\!\cdot\!L)_\nu
\nonumber\\
\!&&\! -\: 8\,\Big( \zeta_\mu\, (\zeta\!\cdot\!L^2)_\nu
\,+\, \zeta_\nu\, (\zeta\!\cdot\!L^2)_\mu\Big)\:
+\:  8\, (\zeta\!\cdot\!L\!\cdot\!\zeta)\, L_{\mu\nu}\:
 -\: 4\,\Big(\zeta_\mu\, (\zeta\!\cdot\!L)_\nu\, +\, \zeta_\nu\, (\zeta\!\cdot\!L)_\mu\Big)\:\widehat{L}\nonumber\\
\!&&\! +\: 4\,\zeta_\mu\zeta_\nu\,\widehat{L}^2\:
-16\,\zeta^2 (L^2)_{\mu\nu}\: +\: 4\,\zeta^2 L_{\mu\nu} \widehat{L}\,\Big]\,,\\[3mm]
\label{eq:gammaM}
(\gamma_M)_\mu \!&\equiv&\! \frac{dM_\mu}{dt}\ =\
-\,\lim_{R\to 0} \bigg( \lim_{\varepsilon \to 0}\,2\varepsilon\: \frac{\partial V^{\rm S,UV}_{\rm eff}[R]}{\partial R^\mu }\bigg)\ =\ 
\frac{1}{16\pi^2}\,\Big[ 24\, (\zeta\!\cdot\! M)\,(\zeta\!\cdot\!L)_\mu\: -\: 4\,(\zeta\!\cdot\!M)\, \zeta_\mu\:\widehat{L}\nonumber\\
\!&&\! -\: 12\,\zeta^2 (M\!\cdot\!L)_\mu\: +\: 2\,\zeta^2 M_\mu\:\widehat{L}\,\Big]\,.
\end{eqnarray}
Fixing $\zeta_\mu$ to its canonical form given in~\eqref{eq:Zmu}, the O(1,3) symmetry breaks to its little group O(3) $\subset$ O(1,3). In this case, $\beta_L$ and $\gamma_M$ go over to five different O(3)-covariant sub-structures. Upon inclusion of a factor of 4 due to a different normalisation of $R^\mu$, the five emergent sub-structures are in perfect agreement with those reported in~\cite{Bednyakov:2018cmx} in the canonical field basis~\eqref{eq:Zmu}.

It is interesting to analyse the UV behaviour of~$V^{\rm S}_{\mathrm{eff}}({\bf \Phi})$ given in~\eqref{eq:VeffScalar}. for a few simple 2HDM scenarios, such as the MS-2HDM~\cite{BhupalDev:2014bir,Darvishi:2019ltl}.
In the MS-2HDM, only the quartic-coupling element $L_{00}$ is non-zero in $L_{\mu\nu}$, whereas $M_\mu$ can be arbitrary, if we assume the soft-breaking of Sp(4) for phenomenological reasons~\cite{BhupalDev:2014bir}. In~this case, \eqref{eq:betaL} and~\eqref{eq:gammaM} become
\begin{eqnarray}
    \label{eq:betaL_MS2HDM}
(\beta_L)_{00} \!&=&\! \frac{1}{16\pi^2}\, \Big( 16\, L^2_{00}\: +\: 24\,{\bf L}\!\cdot\!{\bf L}\: +\: 4\, L_{00}\,\text{tr}\,L^{\rm s}\: +\: 4\,\text{tr}\,(L^{\rm s})^2\, \Big)\,,\\
    \label{eq:gammaM0_MS2HDM}
(\gamma_M)_0 \!&=&\! \frac{1}{16\pi^2}\, \Big( 10\, M_0 L_{00}\: +\: 2\,M_0\,\text{tr}\,L^{\rm s}\: +\: 12\, {\bf M}\!\cdot\!{\bf L} \Big)\,,\\
    \label{eq:gammaMvec_MS2HDM}
\gamma_{\bf M} \!&=&\! \frac{1}{16\pi^2}\, \Big( 12\, M_0\, {\bf L}\: +\: 12\,L^{\rm s}\!\cdot{\bf M}\: +\: 2\, {\bf M}\,\big(L_{00} - \text{tr}\,L^{\rm s}\big) \Big)\,,
\end{eqnarray}
whereas all other $\beta_L$-functions for $L_{\mu\nu}$ vanish.
In~\eqref{eq:betaL_MS2HDM}--\eqref{eq:gammaMvec_MS2HDM}, 
we defined the O(3)-vectors, ${\bf L} \equiv \{L_{0i}\}$ and ${\bf M} \equiv \{M_i\}$, 
and the rank-2 tensor, $L^{\rm s} \equiv \{ L_{ij}\}$, with $i,j=1,2,3$. In addition to ${\bf L}={\bf 0}$ and $L^{\rm s} = 0$,
one may require that the squared mass element~$M_0$ of $M_\mu$ is also zero. As a consequence, \eqref{eq:gammaM0_MS2HDM} implies a vanishing $(\gamma_M)_0$, but a non-zero $\gamma_{\bf M} \propto L_{00} {\bf M}$ [cf.~\eqref{eq:gammaMvec_MS2HDM}]. These results are in agreement with the findings  in~\cite{Ferreira:2023dke}. Nonetheless, the authors 
of~\cite{Ferreira:2023dke} also hinted at the potential existence of additional symmetries in the 2HDM resulting from the possibility $R^0\to -R^0$ in~\eqref{eq:VR} that can be realised in variants of specific globally symmetric scenarios that have been catalogued in~\cite{Ivanov:2007de,Battye:2011jj,Pilaftsis:2011ed}. For instance, one such scenario could be a variation of the so-called CP2 symmetry, with $M_0 = 0$ but $M^2\neq 0$. Notice that this variant is already contained in the MS-2HDM softly broken by a non-zero~$M^2$~\cite{BhupalDev:2014bir}. However, as one could see from~\eqref{eq:n1term} and~\eqref{eq:VeffScalar}, quadratically UV-divergent terms emerge in the effective potential~$V_{\rm eff}$ that are proportional to $L_{00} R^0$, since $\text{tr}\,L = L_{00}$ for the model at hand. In the DR scheme, such terms vanish, but reappear as part of the UV-finite threshold corrections~\cite{BhupalDev:2014bir} in odd powers of $R^0$, e.g.~through terms like $L^{2k+1}_{00} (R^0)^{2k+1}/(M^2)^{2k -1}$, for~${k \ge 1}$.  We therefore find that the additional symmetries observed in the form of RG invariants in~\cite{Ferreira:2023dke}
are violated by UV-finite corrections to $V_{\rm eff}$ at the~one-loop level.

\subsection{Gauge-Boson Loops}\label{subsec:GB}

We will now calculate the contributions from $W^{a=1,2,3}$ and $B$
gauge-boson loops to the effective potential, i.e.~$V^{\rm GB}_{\mathrm{eff}}$. In the $R_\xi$ gauge (with $\xi \to 0^+$), the Hessian of the action describing the gauge system $W^I\equiv (B\,, W^a)$, with $I = 0,a$ and $W^0\equiv B$, is given by 
\begin{eqnarray}
    \label{eq:HessBW}
    H^{\alpha\beta}_{IJ}({\bf \Phi}) \!&=&\! \big(\!-\eta^{\alpha\beta} k^2 + k^\alpha k^\beta \big)\,\delta_{IJ}\: +\: \eta^{\alpha\beta} \Big(\frac{g'^2}{8}\delta_{I0}\delta_{J0} + \frac{g^2}{8} \delta_{Ia}\delta_{Jb}\Big)\,(\zeta\!\cdot\!R)\nonumber\\
 \!&&\! +\: \eta^{\alpha\beta}\,\frac{g g'}{8} \Big(\delta_{I0}\delta_{Ja} + 
 \delta_{Ia}\delta_{J0}\Big)\, \zeta_\mu\, {\bf\bar{\Phi}}\,\Gamma^\mu\Gamma^5 ({\bf 1}_4\otimes \sigma^a)\,{\bf \Phi}\, .
\end{eqnarray}
Hence, the eigenvalues of the $4\times 4$ matrix,
\begin{equation}
  \label{eq:HIJ}
H_{IJ}({\bf \Phi})\ =\ -\frac{1}{d-1}\,\Big(\eta_{\alpha\beta} - \frac{k_\alpha k_\beta}{k^2}\Big)  H^{\alpha\beta}_{IJ}({\bf \Phi})\,,   
\end{equation}
which enter the one-loop effective potential, can be determined using the Cayley--Hamilton polynomial of traces or otherwise~\cite{Cao:2023kgq}. In this regard,  a useful identity proves to be
\begin{equation}
    \label{eq:IdentGB}
\zeta_\mu\zeta_\nu\, \Big({\bf\bar{\Phi}}\,\Gamma^\mu\Gamma^5 ({\bf 1}_4\otimes \sigma^a)\,{\bf \Phi}\Big)\:\Big( {\bf\bar{\Phi}}\,\Gamma^\nu\Gamma^5 ({\bf 1}_4\otimes \sigma^a)\,{\bf \Phi}\Big)\ =\  (\zeta\!\cdot\!R)^2\: -\: \zeta^2\, R^2\, ,   
\end{equation}
which is obtained by employing the identity of the SU(2) generators:  $\sigma^a_{ij} \sigma^a_{kl} = 2\delta_{il}\delta_{kj}\ - \delta_{ij}\delta_{kl}$, in the intermediate steps of the computation. Observe that this identity 
can be deduced from the completeness relation in~\eqref{eq:Complete}. 
Hence, all four eigenvalues of $H_{IJ}$ depend on $R^\mu$ only. 
These are found to be
\begin{eqnarray}
   \label{eq:MWplus}
M^2_{W}[R] \!&=&\!  \frac{g^2}{8}\, (\zeta\!\cdot\!R)\,,\nonumber\\
M^2_{Z}[R] \!&=&\!  \frac{g^2+g'^2}{16}\, \Big( \zeta\!\cdot\!R\: +\: \sqrt{(\zeta\!\cdot\!R)^2\, -\, 4g^2g'^2 \zeta^2 R^2 }\,\Big)\,,\\
M^2_A[R] \!&=&\!  \frac{g^2+g'^2}{16}\, \Big( \zeta\!\cdot\!R\: -\: \sqrt{(\zeta\!\cdot\!R)^2\, -\, 4g^2g'^2 \zeta^2 R^2 }\,\Big)\,.\nonumber
\end{eqnarray}
For field values that go along a neutral flat direction, as could happen in a scale-invariant version of the 2HDM~\cite{Lee:2012jn}, we have  $R^2\equiv R\!\cdot\!R = 0$ and one of the gauge fields, the photon $A$, remains massless. Instead, the other neutral gauge field, associated with the $Z$ boson, will acquire an $R$-dependent mass, similar to the $W^\pm$ bosons. 
Taking the squared-mass expressions~\eqref{eq:MWplus} into account, the effective potential from gauge-boson loops may be cast into the O(1,3)-invariant form:
\begin{eqnarray}
    \label{eq:VeffGB}
 V^{\rm GB}_{\mathrm{eff}}[R]  \!&=&\!  \frac{1}{64\pi^2}\, \bigg[\,
 6 M^4_W [R]\, \bigg( \ln\!\frac{M^2_W[R]}{\bar{\mu}^2}\ -\ \frac{5}{6}\,\bigg)\ +\ 3 M^4_Z [R]\, \bigg( \ln\!\frac{M^2_Z[R]}{\bar{\mu}^2} \ -\ \frac{5}{6}\,\bigg)\nonumber\\
 &&\hspace{1.2cm}+\: 3 M^4_A [R]\, \bigg( \ln\!\frac{M^2_A[R]}{\bar{\mu}^2}\ -\ \frac{5}{6}\,\bigg)\, \bigg]\,.  
\end{eqnarray}
This last result is obtained in the so-called $\overline{\text{MS}}$ scheme (also called the $\overline{\text{DR}}$ scheme), in which the $1/\varepsilon$ UV-poles have been renormalised away. 
In the limit $g' \to 0$, the gauge-boson interactions are invariant
under the maximal custodial symmetry group Sp(4) in the ${\bf \Phi}$-space and as such, they do not affect the form of the MS-2HDM potential. In this limit, the gauge-boson-induced effective potential, $V^{\rm GB}_{\mathrm{eff}}[R]$, becomes a function of $\zeta\!\cdot\!R = R^0$ only. If we turn on the hypercharge gauge coupling, $g'\neq 0$, $V^{\rm GB}_{\mathrm{eff}}[R]$ will depend on both $R^0$ and $R^2$ and the O(1,3)-invariant form of $V^{\rm GB}_{\mathrm{eff}}$ reduces to the little group O(3) in the $R$-space. In this latter limit, our results agree with those reported earlier in~\cite{Cao:2022rgh,Cao:2023kgq}.

\subsection{Fermion Loops}\label{subsec:F}

The Dirac-algebra formalism that we have been studying here can also be applied successfully to include fermion loops in the effective potential in an O(1,3)-invariant manner. For this purpose, let us consider a single family of $u$- and $d$-quarks, with $Q = (u\,, d)^{\sf T}$. The generalisation to more families is straightforward, but it will be given elsewhere. 

The Hessian of the combined $Q$-system derived from the Yukawa Lagrangian~${\cal L}_{\rm Y}$ in~\eqref{eq:LYuk} is given by 
\begin{equation}
   \label{eq:HessQ}
H_Q ({\bf \Phi})\ =\ {\bf 1}_2\slashed{k} \: -\: {\cal M}_Q({\bf \Phi})\, P_R\: -\:  {\cal M}^\dagger_Q({\bf \Phi})\, P_L\;,
\end{equation}
where $P_{L\,(R)} \equiv [{\bf 1}_4 -(+) \gamma_5]/2$ are the chirality projection operators for the Dirac $u$- and $d$-quarks, such that $Q_{L(R)} = P_{L(R)}\,Q$. Furthermore, ${\cal M}_Q({\bf \Phi})$ is a gauge-covariant ${\bf \Phi}$-dependent quark-mass matrix given in~\eqref{eq:calMQ}. For a single family of quarks, ${\cal M}_Q({\bf \Phi})$ is simply a $2\times 2$-dimensional matrix, which becomes $6\times 6$-dimensional when all three quark families are included.

In order to compute the fermionic loop contribution to the effective potential, we use the formula~\eqref{eq:V1loopk}.
In this way, we find in the $\overline{\text{DR}}$ scheme,
\begin{equation}
   \label{eq:V1loopF}
 V^{\rm F}_{\mathrm{eff}}({\bf \Phi})\ =\ 
 \frac{4 N_c}{64\pi^2}\: {\rm tr}\,\bigg[\,
 ({\cal M}^\dagger_Q \, {\cal M}_Q )^2 \, \bigg(\ln\!\frac{{\cal M}^\dagger_Q \, {\cal M}_Q}{\bar{\mu}^2}\ -\ 1\,\bigg)\,\bigg]\; , \end{equation}
where $N_c = 3$ are the quark colours  and
\begin{equation}
   \label{eq:MQMQ}
{\cal M}^\dagger_Q \, {\cal M}_Q \ =\  \left(\! \begin{array}{cc}
 h^{u*}_i h^u_j\, (\phi^\dagger_j\phi_i)    &  -\,h^{u*}_i h^d_j\, (\phi^{\sf T}_i \,i\sigma^2 \phi_j)\\
h^{d*}_k h^u_l\, (\phi^\dagger_k\,i\sigma^2 \phi^*_l)  & h^{d*}_k h^d_l\, (\phi^\dagger_k\phi_l)
\end{array}
\!\right)\, ,  
\end{equation}
with $i,j,k,l = 1,2$ being the summation indices that run over the two scalar doublets, $\phi_{1,2}$. By~employing the completeness relation~\eqref{eq:Complete}, we may turn the diagonal SM-gauge-invariant elements of the matrix in~\eqref{eq:MQMQ} into the O(1,3)-covariant expressions,
\begin{eqnarray}
    \label{eq:HuHd}
h^{u*}_i h^u_j\, (\phi^\dagger_j\phi_i)\: =\: \frac{1}{4}\, Y^u_\mu\, R^\mu \,,\qquad  
h^{d*}_k h^d_l\, (\phi^\dagger_k\phi_l)\: =\: \frac{1}{4}\, Y^d_\mu\, R^\mu  \, ,
\end{eqnarray}
where the following Yukawa four-vectors have been introduced~\cite{Cao:2022rgh}: $Y^u_\mu \equiv h^{u\,\dagger} \bar{\sigma}_\mu\, h^u$ and 
$Y^d_\mu \equiv h^{d\,{\sf T}} \bar{\sigma}_\mu\, h^{d\,*}$, in the vector representation: $h^{u\,(d)} = (h^{u\,(d)}_1,\, h^{u\,(d)}_2)^{\sf T}$. It is not difficult to convince ourselves that the Yukawa four-vector $Y^{u\,(d)}_\mu$ is real and has null norm, i.e.~$Y^{u\,(d)}_\mu\, =\, Y^{u\,(d)\,*}_\mu$ and 
$Y^{u\,(d)}\cdot Y^{u\,(d)} = 0$, although one has in general: $Y^{u}\!\cdot\!Y^{d} \neq 0$. Moreover, when calculating the determinant of the two-by-two matrix, ${\cal M}^\dagger_Q \, {\cal M}_Q$, we have to translate its absolute squared element, $|({\cal M}^\dagger_Q \, {\cal M}_Q)_{12}|^2$, into a manifestly U(1)$_Y$-invariant expression, i.e.\footnote[3]{To arrive at the last formula in~\eqref{eq:Hud}, we use the known identity: $\varepsilon_{\alpha\beta}\, \varepsilon_{\gamma\delta} = \delta_{\alpha\gamma}\,\delta_{\beta\delta} - \delta_{\alpha\delta}\,\delta_{\beta\gamma}$, where the indices $\alpha,\beta,\gamma,\delta =1,2$ take values in the SU(2)$_L$-space for each of the two scalar doublets~$\phi_{1,2}$.}
\begin{eqnarray}
    \label{eq:Hud}
 |({\cal M}^\dagger_Q \, {\cal M}_Q)_{12}|^2\: =\: -\,h^{u*}_i h^d_j h^{d*}_k h^u_l\, (\phi^{\sf T}_i \,i\sigma^2 \phi_j)
 \, (\phi^\dagger_k\,i\sigma^2 \phi^*_l)
 \: =\:  
\frac{1}{16}\, \Big( Y^{du}_\mu\,Y^{du\,*}_\nu -\, Y^d_\mu\,Y^u_\nu\Big)\, R^\mu R^\nu  ,\ \quad
\end{eqnarray}
where $Y^{du}_\mu \equiv h^{d\,{\sf T}} \bar{\sigma}_\mu\, h^u$ is a complex O(1,3) vector which obeys the null properties:\footnote[4]{To prove these properties, one may use the identity: $(\bar{\sigma}_\mu)_{ij}(\bar{\sigma}^\mu)_{kl} = 2\,(\delta_{ij}\delta_{kl} - \delta_{il}\delta_{kj})$.}
\begin{equation}
  \label{eq:Ynull}
 Y^{du}\!\cdot\! Y^{du}\: =\: Y^{du}\!\cdot\! Y^{d}\: =\: Y^{du}\!\cdot\! Y^{u} \: =\: 0\, .   
\end{equation}
Making now use of~\eqref{eq:HuHd}
and~\eqref{eq:Hud}, the two eigenvalues of ${\cal M}^\dagger_Q \, {\cal M}_Q$ acquire an O(1,3)-invariant form:
\begin{equation}
    \label{eq:m2ud}
m^2_{u\,(d)}[R]\ =\ \frac{1}{8}\, \Big[\, Y^u\!\cdot\!R\, +\, 
Y^d\!\cdot \!R\ 
+(-)\ \sqrt{\big(Y^u\!\cdot\!R\, +\, Y^d\!\cdot\!R\big)^2 -\: 4\, \big|Y^{du}\!\cdot\!R\big|^2}\:
\Big]\,.
\end{equation}
This formula is consistent with a related gauge-invariant expression that was derived before in~\cite{Carena:2000yi} for the Type-II 2HDM potential of the CP-violating MSSM~\cite{Pilaftsis:1999qt}. For instance, along neutral field directions (for which $R^2 = 0$), the RHS of~\eqref{eq:Hud} vanishes identically, and the two eigenvalues in~\eqref{eq:m2ud}
reduce to: $m^2_{u,d}[R]\ =\ \frac{1}{4}\, Y^{u,d}\!\cdot\!R$, as they should be. Likewise, by setting $h^d = h^{u\,*}$, we have the usual custodial symmetric case, for which $Y^{du}_\mu = Y^{d}_\mu = Y^{u}_\mu$, yielding the expected equality between the $u$- and $d$-quark masses: 
$m^2_u[R] = m^2_d [R]$. We must remark here that the analytic expressions for the $u$- and $d$-quark masses in~\eqref{eq:m2ud} differ from those presented in~\cite{Cao:2022rgh,Cao:2023kgq}, through the introduction of a new Yukawa four-vector:~$Y^{du}_\mu$.

Given the analytic results of the $R$-dependent squared mass eigenvalues in~\eqref{eq:m2ud}, the fermionic one-loop contribution to the effective potential becomes
\begin{equation}
   \label{eq:V1FR}
 V^{\rm F}_{\mathrm{eff}}[R]\ =\ 
 \frac{N_c}{16\pi^2}\, \bigg[\,
 m^4_u[R] \, \bigg( \ln\!\frac{m^2_u[R]}{\bar{\mu}^2}\: -\: 1\,\bigg)\ +\ m^4_d[R] \, \bigg( \ln\!\frac{m^2_d[R]}{\bar{\mu}^2}\: -\: 1\,\bigg)\bigg]\; , 
\end{equation}
which is manifestly invariant under O(1,3) reparameterisations in the $R$-space. A general quark-Yukawa sector breaks all 13 global symmetries of the 2HDM potential tabulated in~\cite{Pilaftsis:2011ed} at the one-loop level. But, one may envisage grand unification scenarios for the third family Yukawa couplings, with~$h^t = h^b$ at large $\tan\beta \sim 40$, leading to: $Y^{tb}_\mu = Y^{t}_\mu = Y^{b}_\mu$, thus preserving some of the custodial symmetries~\footnote[5]{These symmetries need to be  violated subdominantly by the Yukawa couplings of the lighter quarks, so as to allow for a realistic realisation of the observed quark-mass spectrum.}. Nevertheless, the maximal custodial symmetry Sp(4) of the MS-2HDM will always be broken by $u$- and $d$-quark loops, since Sp(4) can only be preserved if $Y^{tb}_\mu\,,\, Y^{t}_\mu\,,\, Y^{b}_\mu\, \propto\, \zeta_\mu$. This can never be the case, because $\zeta^2 \neq 0$, whilst $Y^{tb}_\mu\,,\, Y^{t}_\mu\,,\, Y^{b}_\mu$ have all null norms [cf.~\eqref{eq:Ynull}]. 

From the above considerations, it becomes evident that an interesting SO(1,3) reparameter\-isation invariant quantity characterising geometric flavour misalignments in the Yukawa sector is given by the complex-valued 4D pseudo-scalar,
\begin{equation}
    \label{eq:MisYuk}
Y_F\ =\  \varepsilon^{\mu\nu\rho\sigma}\, \zeta_\mu\, Y^{du}_\nu\, Y^d_\rho\,  Y^u_\sigma\, .   
\end{equation}
A detailed study of the role of $Y_F$ in 2HDM scenarios will be given elsewhere.

\section{Conclusions}\label{sec:concl}

We have presented a novel formulation of the covariant bilinear formalism for the Two Higgs Doublet Model, which is based on 
the Dirac algebra associated with the $\text{SL}(2,\mathbb{C})$ group that acts on the scalar doublet field 
space~[cf.~\eqref{eq:Dirac}]. 
This new Dirac-algebra formulation has allowed us to succinctly express the one-loop effective potential $V_{\rm eff}$ of a general 2HDM in a fully covariant form in the bilinear field space. In particular, $V_{\rm eff}$ is invariant under O(1,3) field reparameterisations in the bilinear field space, which was called the $R$-space. To achieve this O(1,3)-covariance in the $R$-space, we must also consider the wave-function four-vector $\zeta_\mu$ that arises from the kinetic terms of the two scalar doublets. Fixing $\zeta_\mu$ to its canonical form in~\eqref{eq:Zmu} breaks O(1,3) to its little group O(3), which was the working hypothesis in all previous studies at the quantum loop level~\cite{Bednyakov:2018cmx,Ferreira:2023dke,Cao:2022rgh,Cao:2023kgq}. 

In the present Dirac-algebra formulation, the two scalar doublets $\phi_{1,2}$, together with their hypercharge conjugates, form an 8D multiplet, called the ${\bf \Phi}$-multiplet in~\eqref{eq:Phi}. 
The so-constructed ${\bf \Phi}$-multiplet has transformation properties akin to a Majorana fermion, exhibiting an analogous left- and right- `chiral' decomposition in terms of the U(1)$_Y$ hypercharge group of the SM. An important outcome of this framework is the analytic result given in~\eqref{eq:VeffScalar}, which provides a manifestly O(1,3)-invariant and IR-safe expression for the one-loop effective potential from scalar loops in the bilinear $R$-space. In particular, the IR poles due to the loop-momentum expansion of~$1/(k^2)^n$ (with $n\ge 2$) that occur in the O(3)-covariant effective-potential computation of~\cite{Cao:2023kgq} can be avoided.

We have elucidated by virtue of a few simple examples how the Dirac-algebra formalism can be used to evaluate the breaking of global symmetries of the 2HDM potential by scalar, gauge-boson and fermion loops, in a field-reparameterisation invariant manner. We find that no new symmetries (which may manifest themselves in the form of RG invariants) survive in the 2HDM effective potential, once all UV-finite and IR-safe corrections to the latter are computed. It would be of great interest to analyse in detail the breaking patterns of the 13 global symmetries of the tree-level 2HDM potential by quantum effects at one- and higher loops, using the Dirac-algebra formalism presented in this work.

\subsection*{Acknowledgements} I would like to thank Bohdan Grzadkowski for discussions regarding Ref.~\cite{Ferreira:2023dke}.
This work is supported in part by
the STFC research grant: ST/X00077X/1.

\vfill\eject
\bibliography{Dirac}

\end{document}